\documentclass[conference]{IEEEtran}
\IEEEoverridecommandlockouts
\usepackage{cite}
\usepackage{enumitem}
\pdfoutput=1
\usepackage{amsmath,amssymb,amsfonts}
\usepackage{algorithmic}
\usepackage{multirow}
\usepackage{graphicx}
\usepackage{textcomp}
\usepackage{xcolor}
\def\BibTeX{{\rm B\kern-.05em{\sc i\kern-.025em b}\kern-.08em
    T\kern-.1667em\lower.7ex\hbox{E}\kern-.125emX}}
\begin{document}

\title{RO-SVD: A Reconfigurable Hardware Copyright Protection Framework for AIGC Applications
\thanks{*Corresponding author: Zhuoheng Ran (zhuoheran2-c@my.cityu.edu.hk).}
}

\author{\IEEEauthorblockN{Zhuoheng Ran, Muhammad A.A. Abdelgawad, Zekai Zhang, Ray C.C. Cheung and Hong Yan}
\IEEEauthorblockA{Department of Electrical Engineering and Centre for Intelligent Multidimensional Data Analysis, \\City University of Hong Kong, Hong Kong SAR, China\\
Email: \{zhuoheran2-c, mabdelgaw2-c, zekazhang2-c\}@my.cityu.edu.hk, \{ray.cheung, h.yan\}@cityu.edu.hk}
}

\maketitle

\begin{abstract}
The dramatic surge in the utilisation of generative artificial intelligence (GenAI) underscores the need for a secure and efficient mechanism to responsibly manage, use and disseminate multi-dimensional data generated by artificial intelligence (AI). In this paper, we propose a blockchain-based copyright traceability framework called ring oscillator-singular value decomposition (RO-SVD), which introduces decomposition computing to approximate low-rank matrices generated from hardware entropy sources and establishes an AI-generated content (AIGC) copyright traceability mechanism at the device level. By leveraging the parallelism and reconfigurability of field-programmable gate arrays (FPGAs), our framework can be easily constructed on existing AI-accelerated devices and provide a low-cost solution to emerging copyright issues of AIGC. We developed a hardware-software (HW/SW) co-design prototype based on comprehensive analysis and on-board experiments with multiple AI-applicable FPGAs. Using AI-generated images as a case study, our framework demonstrated effectiveness and emphasised customisation, unpredictability, efficiency, management and reconfigurability. To the best of our knowledge, this is the first practical hardware study discussing and implementing copyright traceability specifically for AI-generated content.
\end{abstract}

\begin{IEEEkeywords}
AI-generated content, copyright protection, singular value decomposition (SVD), blockchain, nanofabrication, AI security, low-power AI
\end{IEEEkeywords}

\begin{figure*}[t!]
\centering
\includegraphics[width=1\linewidth]{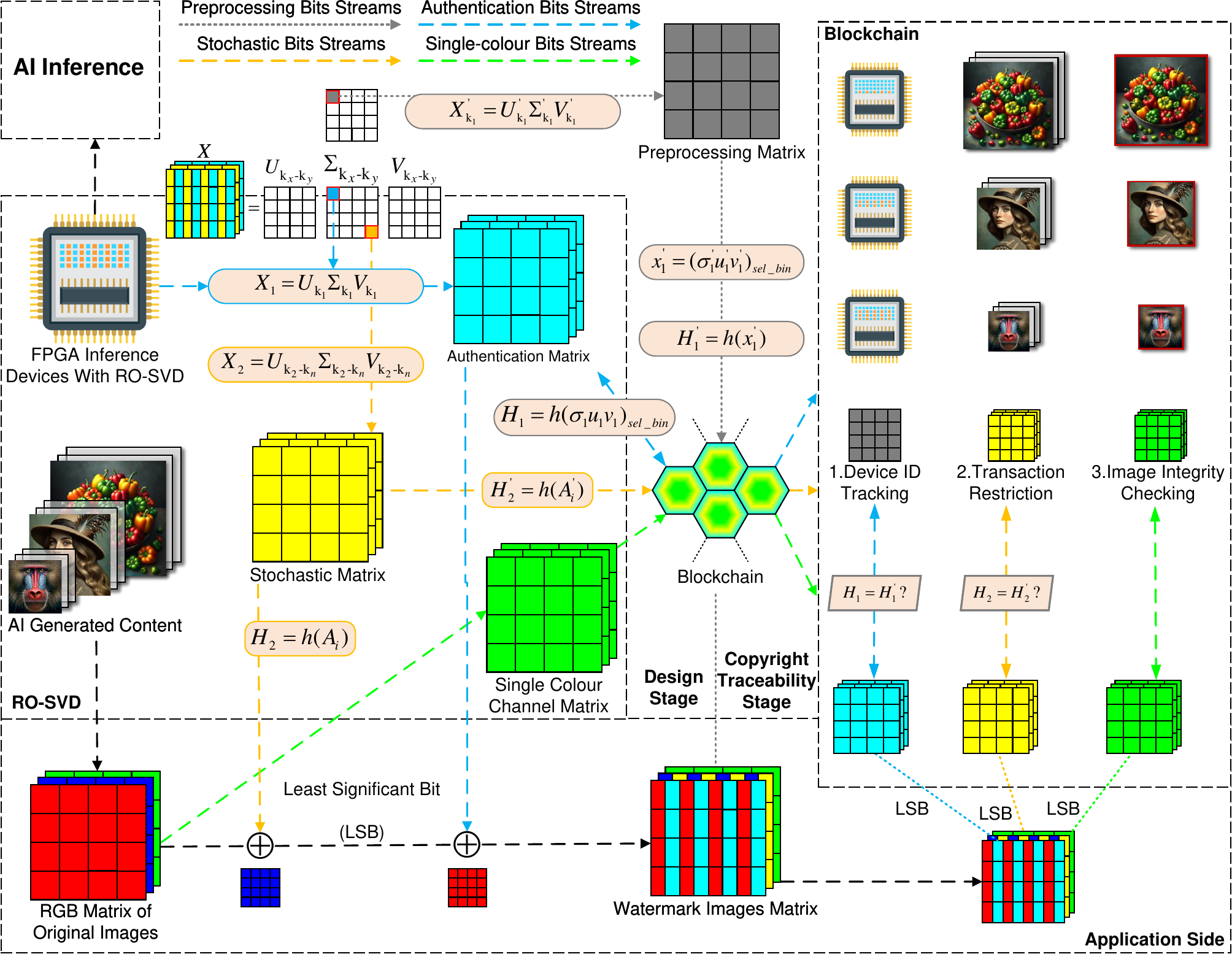}
\caption{The overall flow of the proposed framework with the least significant bit (LSB) lossless application side.}
\label{Overall.Process}
\end{figure*}

\section{Introduction}
As generative artificial intelligence (GenAI) becomes increasingly prevalent, notable products such as ChatGPT, Midjourney and Sora have spurred reflections on the evolving interactions between humans and artificial intelligence (AI). This paradigm shift acts as a double-edged sword, fostering significant growth in creative industries while simultaneously raising widespread public concerns about the copyright issues of AI-generated content (AIGC)\cite{lee2023survey}, thereby hindering the development of these technologies to some extent. The creation process of AI content involves three components that allow personalised content generation: data, algorithms and accelerated computing\cite{xu2024unleashing}. The field-programmable gate array (FPGA) is regarded as a reconfigurable and low-power solution for AI computing\cite{han2023hpta,dong2023heatvit,li2023unified,nag2023vita,wang2022via}, supporting efficient computing while providing an additional hardware layer to deploy security functions in parallel with inference operations.

Known for distributed ledger technology (DLT), blockchain technology is extensively covered in recent literature as a strategy for AIGC copyright protection due to its efficient management and transaction capabilities for digital copyright\cite{chen2023challenges, liu2024blockchain,chen2023towards}. Additionally, efforts focused on hardware variability in FPGA devices can be exploited by utilising multiple groups of ring oscillators (ROs) configured to flexibly construct lightweight security primitives based on the delay of silicon, constituting a critical component for conventional purposes such as intellectual property (IP) protection, digital fingerprinting and encryption for Internet of Things (IoT) applications\cite{cirne2024hardware,mao2023realise,plusquellic2023privacy}. By exploiting the unpredictability of hardware entropy sources derived from FPGA variability in conjunction with blockchain technology, each digital asset can be assigned a unique and unclonable identity that is traceable within the blockchain and IoT ecosystems. However, few practical implementations are specifically designed for emerging AIGC copyright protection issues due to novel engineering challenges.

By targeting AI-generated images as a case study, our work is expected to address practical engineering challenges in efficiency, cost, management, unpredictability and reconfigurability for emerging AI-generated content copyright traceability issues. Our main contributions are as follows:
\begin{enumerate}[label={(\arabic*)}]
\item We demonstrated that matrix-based responses from hardware entropy sources exhibit approximate low-rank properties due to process variations, offering the potential for AIGC copyright traceability at the device level.
\item On this basis, we proposed a framework called ring oscillator-singular value decomposition (RO-SVD), which enables automatic and efficient marking and management of AIGC based on blockchain.
\item In the proposed framework, we first introduced singular value decomposition (SVD) as a robust computing component to separate matrix-based primitives generated by FPGAs and act as physically unclonable functions (PUFs) and true random number generators (TRNGs) co-designs for management and marking massive content by a reconfigurable and unpredictable way.
\item We developed a hardware-software (HW/SW) co-design prototype that embodies the principles of our work and can be easily constructed on existing devices.
\end{enumerate}
To the best of our knowledge, this is the first practical hardware engineering study discussing and implementing copyright traceability specifically for AI-generated content. Furthermore, our work paves the way for copyright traceability of AI-generated content with multiple dimensions and forms.

\section{Related Work and Background}
Robust intellectual property protection schemes are essential for the legitimate use of AI-generated content. Existing schemes like metadata tracking can be easily modified and are typically combined with watermarking techniques\cite{martynovich2024toward}. For example, Stable Diffusion uses DCT-DWT-SVD to distinguish machine-generated images\cite{stablediffusion}. Cryptographic algorithms like homomorphic hashing utilise homomorphic properties to enable content integrity checking without comparing the original data\cite{cryptoeprint:2024/498}. Blockchain technology is a valuable reference for AIGC copyright management. It can provide a clear and unalterable record of copyright and creation history from creation time to copyright information, which can be encrypted and stored on the blockchain-based framework\cite{sanka2021systematic}. The blockchain can also automatically execute copyright-related transactions and agreements using smart contracts\cite{sanka2021survey}, which can increase the transparency and efficiency of transactions and reduce the risk of copyright infringement and disputes.

Process variation in digital circuits is widely used for emerging security functions\cite{liu2016bias}. A dataset for frequency in 217 FPGAs from the University of Belfast at HOST\cite{hesselbarth2018large} provides further evidence that the physical layout function of spatial variation can be established as a component of intra-die and has potential for optimisation in 28nm Xilinx FPGAs. Literature\cite{asha2021improving} further constrains the direction of principal components along the Y-coordinates by setting the $\{Q_{y_i}\}$. However, optimising layouts to produce reliable responses for lightweight IoT applications makes it difficult to meet the reconfigurable requirements in AIGC applications. In addition to process variations, the output is also affected by timing effects\cite{karimi2021security} or environmental variations\cite{song2021environmental}. However, existing schemes on the generation side are primarily designed to be simple enough to avoid costs and rely on decision algorithms for IoT applications\cite{ali2022secure,streit2021design,rullo2024puf}. Due to the enormous amount of AI-generated content, the original response should also be assumed to be proportional and massive, which makes it difficult for decision-making algorithms to accurately describe the characterisation of large amounts of data generated by entropy sources. Secure sketching schemes like Code-Offset and Syndrome Construction typically serve as an integrated method, which stores helper data for consistent seed generation but potentially exposes response patterns\cite{gu2020modeling,qureshi2021puf,lounis2023lessons}.

\section{Computational Structure}
We introduce singular value decomposition (SVD) to decompose original matrices generated from the entropy source, extracting singular values and corresponding vectors for authentication and stochastic purposes. Processed bitstreams are then encrypted through a hash function for registration, authentication and other processes, as shown in Figure \ref{Overall.Process}.
\subsection{Overall Procedure of Data Streams}
{\textbf{Pre-processing Bits Streams:}}
FPGA-accelerated devices with the proposed IP component generate an initial matrix and act as a pre-processing procedure for blockchain registration. To achieve this, the first specific number of singular values of $\Sigma_{k_1}'$ and its corresponding left and right singular matrices $U_{k_1}'$ and $V_{k_1}'$ are extracted and separated as a compared authentication matrix by TSVD processing, and the principal component vector is selected and binarised from the reconstructed rank one matrix $X_1'$ is stored as the hash value $H_1'$ in the blockchain for verification in the following verification steps. Grey elements in Figure \ref{Overall.Process} indicate the steps of the pre-processing stage, which is computed only when the RO-SVD is to be reconstructed and updated to maintain its effectiveness. This pre-processing consumes resources only when RO-SVD needs reconstruction during updates.

{\textbf{Authentication and Stochastic Bits Streams:}}
The original generator matrices from entropy source components after SVD decomposition can be simultaneously extracted singular values with corresponding left and right singular value vectors from $k_1$ and $k_2$ to $k_n$ to respectively generate two matrices for authentication and stochastic matrices by TSVD and matrix reconstruction. In the least significant bit (LSB) scenario, the reconstructed authentication matrix $X_1$ and random matrix $X_2$ can be separately hidden in a single channel, and the remaining channel can be retained to verify the integrity of the image. In addition, extracting the authentication matrix $X_1$ and random matrix $X_2$ is straightforward on the copyright traceability. The authentication sequences $H_1$ and $H_1'$ are used to find which device produces generative AI content. Meanwhile, a group of $H_2$ and $H_2'$ is used to limit the number of transactions on the blockchain or other potential uses. Although matrix decomposition computation is resource-intensive, considerable computational consumption shifts to the traceability side and is available when needed to save resources.

{\textbf{Single-colour Bits Streams: }}
Taking the LSB as the application side, the image integrity can be further verified after finding the device image ID on the record. Copyright traceability is achieved by storing the content of the monochrome channel and the associated hash value at the device level.

\subsection{Proposed Methods}
Consider a raw matrix  $\mathbf{A}\in \mathbb{R}^{m \times n}$ generated from the entropy source, which contains the intrinsic feature $\mathbf{A}$ and the stochastic feature $\mathbf{R}$ and can be represented as:
\begin{equation}
\mathbf{A} = \mathbf{X} + \mathbf{R}
\end{equation}

To enable the simultaneous generation of stochastic seed matrices and authenticated seed matrices, the intrinsic feature should be separated from the original matrices and considered an authentication matrix. We observed that the same chips tend to generate similar response matrices $\mathbf{A}_j$ with almost identical patterns due to intra-die variations in entropy sources, as demonstrated in Figure \ref{Correlation1.1} in the experiment. 

We propose to use the singular value decomposition to extract the intrinsic features from the response matrices $\mathbf{A}_j$ and use a group of matrixes $\mathbf{X}_j$ with deficient rank for authentication. Therefore, the estimated principal component matrix $\hat{\mathbf{X}_j}$ of interest for authentication purposes should satisfy:
\begin{equation}
\| \mathbf{A}_j - \hat{\mathbf{X}_j} \|^2_F = j\tau^2
\end{equation}

where $\| \cdot \|_F$ is the Frobenius norm\footnote{For the matrix $C$ with the size of $m \times n$ and elements $c_{ij} (i = 1, \ldots, m, j = 1, \ldots, n)$, the Frobenius norm calculates the square root of the sum of the squares of the differences of the elements of two matrices, i.e., $\|C\|_F = \sqrt{\sum_{i=1}^{m} \sum_{j=1}^{n} |a_{ij}|^2}$.} and $\tau$ is the standard deviation of noise. 

To extract the main pattern $\mathbf{X}$ for authentication purposes from the original matrix $\mathbf{A}$, we used the SVD to get the best low-rank approximation. Assume that the SVD of the original matrix $\mathbf{A}$ is given by
\begin{equation}
\mathbf{A} = \mathbf{U} \mathbf{\Sigma} \mathbf{V}^T
\end{equation}
where \( \mathbf{U} \in \mathbb{R}^{m \times m} \) is an orthogonal matrix whose columns are the left singular vectors. \( \mathbf{\Sigma}\in \mathbb{R}^{m \times n} \) is a diagonal matrix with singular values on its diagonal, ordered in descending manner. \( \mathbf{V}\in \mathbb{R}^{n \times n} \) is another orthogonal matrix whose columns are the right singular vectors. 
The best rank-$k$ approximation of $\mathbf{A}$ is denoted as $\mathbf{A}_k$ and computed by truncated SVD  (TSVD). In other words, $\mathbf{A}_k$ is computing by keeping only the largest $k$ values and their singular vectors, as follows 
\begin{equation}
\mathbf{A}_k  = \mathbf{U} \mathbf{\Sigma}_k \mathbf{V}^T
\end{equation}

where $\mathbf{\Sigma}_k$ is obtained from the matrix \(\Sigma\) by setting the diagonal elements to zeros but the first \(k\) singular values
\begin{equation}
\mathbf{\Sigma}_{k} = \texttt{diag}(\sigma_1, \ldots, \sigma_k, 0, \ldots, 0).
\end{equation}

The random matrix can be generated by keeping the $k$ smallest singular values of $\mathbf{\Sigma}$, as follows 
\begin{equation}
\mathbf{\Sigma}_{\bar{k}} = \texttt{diag}(0, \ldots, 0,\sigma_{n-k}, \ldots,\sigma_{n-2},\sigma_{n-1} ,\sigma_n).
\end{equation}
Therefore, 
\begin{equation}
\mathbf{A}_{\bar{k}}  = \mathbf{U} \mathbf{\Sigma}_{\bar{k}} \mathbf{V}^T
\end{equation}
%Let \( A \) be the matrix obtained after applying the above method. 
First, we compute the row averages:
\[
\bar{\mathbf{a}}_{i} = \frac{1}{m} \sum_{j=1}^{n} {(\mathbf{A}_{\bar{k}})}_{ij}
\]
where \( \bar{\mathbf{a}}_{i} \) is the average of row \( i \), and \( n \) is the number of columns in \( \mathbf{A}_{\bar{k}} \).
Next, we apply the threshold  to each element of \( \mathbf{A}_{\bar{k}} \), comparing it to its row average:
\[
(\mathbf{A}_{\bar{k}})_{ij} = \left\{
    \begin{array}{cl}
        1 & \text{if } (\mathbf{A}_{\bar{k}})_{ij} \geq  \bar{\mathbf{a}}_{i} \\
        0 & \text{otherwise}
    \end{array}
\right.
\]
for \( i = 1, \ldots, m \) and \( j = 1, \ldots, n \), where \( m \) is the number of rows in \( A \). After obtaining a binary random matrix, we apply a hash function to get a hash sequence to make the data stored shorter and less predictable:
\[
\mathbf{H} = \texttt{Hash}(\mathbf{A}_{\bar{k}})
\]
where \( \mathbf{H}  \) is the hash output. Moreover, we consider the $\mathbf{A}_k$  as the authentication sequence, where \( k < \min(m, n) \).

\begin{figure}[t!]
\centering
\includegraphics[width=0.9\linewidth]{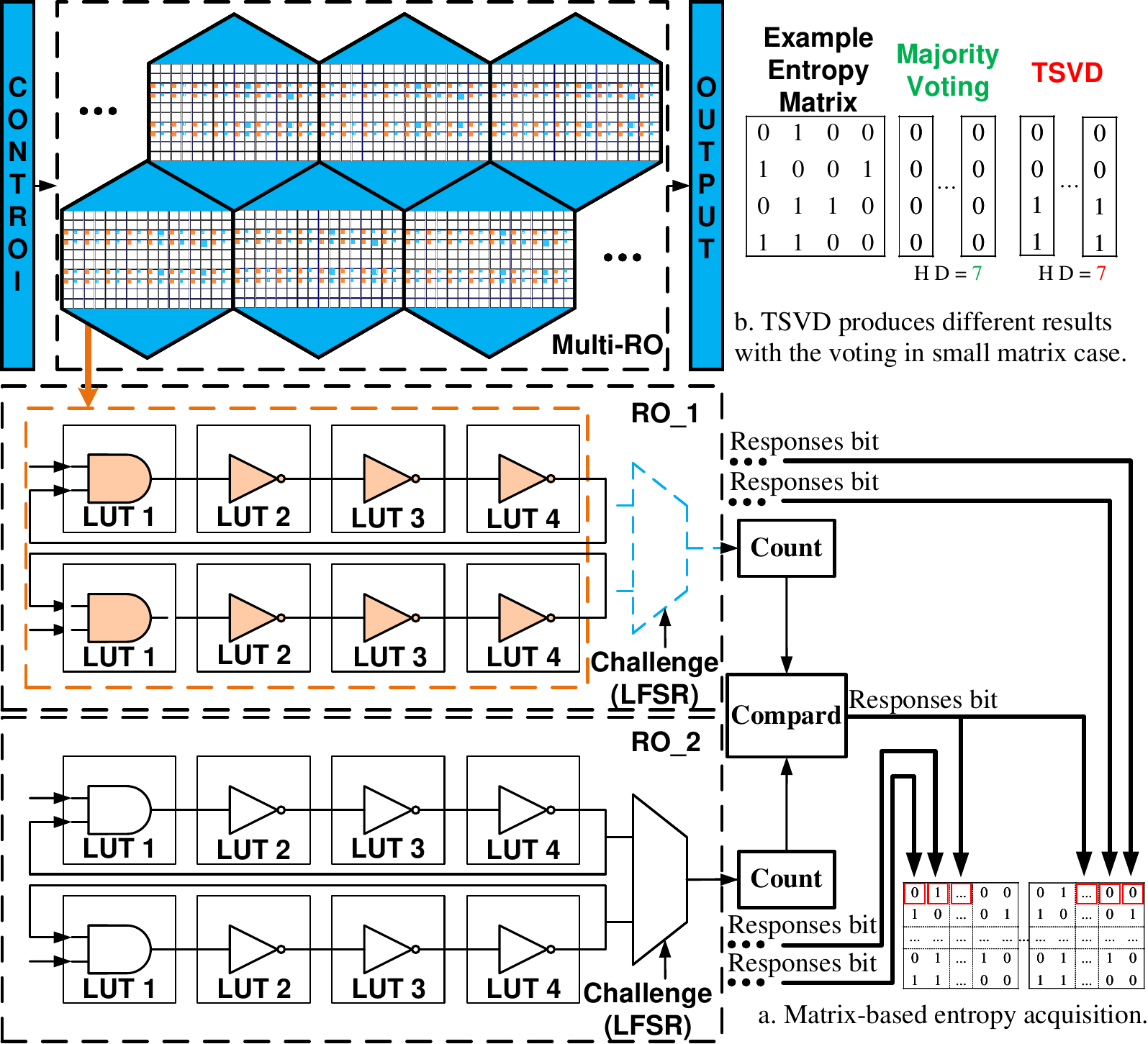}
\caption{Operations and principles of proposed entropy source design.}
\label{Entropy Source}
\end{figure}

\section{Hardware Implementation}
\subsection{Entropy Source Design}
Ring oscillator-based entropy sources in FPGAs can be designed as true random number generators (TRNG) or physical unclonable functions (PUF) for authentication by different considerations. TRNGs generate high entropy and unpredictable random numbers, ensuring the sequence is neither repeatable nor predictable. In contrast, PUFs exploit hardware uniqueness to create device-specific identifiers to ensure output repeatability under the same conditions. TRNGs have low stability requirements and can leverage environmental variations (e.g., temperature) to increase randomness, but PUFs require high stability against environmental variations. TRNGs involve complex post-processing to ensure high-quality randomness such as de-correlation and entropy boosting, while PUFs focus on ensuring consistency and reliability. Therefore, TRNGs and PUFs require different constraint settings, including routing, timing, environmental, and I/O port constraints. Output processing involves helper data algorithms (HDAs) and error correction to ensure consistent responses despite environmental changes or device ageing. Our design reduces constraint settings compared to other delay-based entropy sources, using only a few necessary constraints. Without requiring stable responses, we focus on processing matrix-based raw data from entropy sources to extract authentication and stochastic seeds using SVD components, as shown in Figure \ref{Entropy Source} (a). This new architecture allows marking massive content using unstable entropy and provides a flexible renewal mechanism.

\subsection{Singular Value Decomposition}
We improved the design \cite{abdelgawad2022high} from FPL to build the singular value decomposition (SVD) hardware module using Vivado High-Level Synthesis (HLS) based on the Jacobi method. We integrated the adapted AXI protocol and input/output matrices in floating-point format into our framework. The AXI protocol significantly improves the communication efficiency and flexibility between the SVD module and external memory. Although our framework is low-precision sensitive, converting the input and output matrices to floating-point accommodates high-precision values. This transition is critical for pattern recognition if larger matrices are required. These enhancements pose challenges in computational complexity and resource utilisation, but our design minimises latency and size impact while meeting the requirements.

\subsection{Interface Design}
We created a new IP core based on the lightweight AXI4-Lite protocol that enables parallel operations by customising programmable logic (PL) as the master device in AXI protocol-based communications in Zynq 7000 SoC for our work. Our design allows the exchange of control signals and data streams with DDR RAM within millisecond-level delays through DMA, improving performance and reducing the CPU workload. We used the high-performance AXI HP interface instead of AXI GP for data transfer tasks that can provide twice the bandwidth and direct access to DDR RAM. The HP interface is only available if the PL serves as the master device. Multiple lightweight entropy sources are connected in parallel to the interface, trading a small amount of hardware size for higher throughput in a single cycle. This allows the hardware interface to be synthesised into registers and LUTRAMs instead of BRAMs by CAD tools. Registers provide high-speed synchronous data read/write operations for data buffers, and LUTRAMs enable immediate asynchronous read operations to control signal transmission.

\begin{figure*}[t]
\centering
\includegraphics[width=1\linewidth]{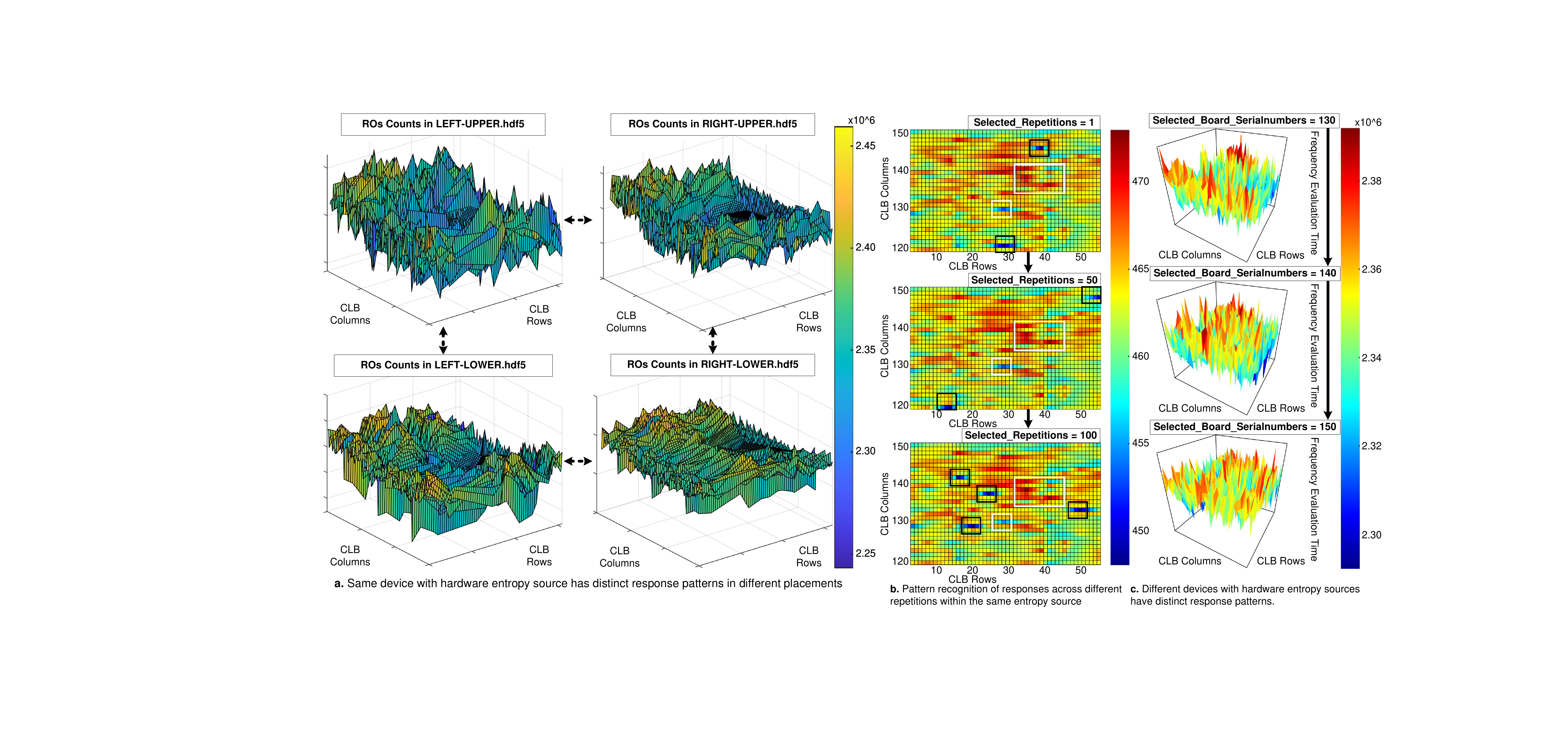}
\caption{Pattern recognition of hardware entropy sources.}
\label{Pattern Analysis}
\end{figure*}

\subsection{Application on Blockchain}
\subsubsection{Lossless Case-Non-Fungible Token (NFT)}
Ownership can be declared with a lossless format in the Non-Fungible Token (NFT). Each unique matrix can be embedded in the lossless images by LSB, which does not affect the visual effect but ensures association with images. We can declare the FPGA board (unique ID) as the author of generated images to enable copyright traceability at the device level while enhancing NFT uniqueness. The generated random matrix limits transactions and the distribution of images to prevent speculation.

\subsubsection{Robust Case-DCT-DWT-SVD}
Our work can be dynamic binary watermark images in the DCT-DWT-SVD, which minimises the impact on image quality and significantly reduces image quality loss. The remaining random matrix can also limit the number of transactions, which helps write smart contract code that defines image usage rights and transaction rules, such as automating copyright transfers.

\section{Results}
\subsection{Experimental Setup}
The hardware experiment uses hundreds of PYNQ-Z2 boards featuring the ZYNQ XC7Z020-1CLG400C SoC, which is a preferred choice for AI-accelerated development and integration due to its compatibility with Python for rapid AI prototyping and extensive library support. The design is implemented using VHDL and High-Level Synthesis (HLS) with Vivado (2023.1), Vitis (2023.1) and Vitis HLS (2023.1). The analysis and experimentation software includes MATLAB (R2023b), RStudio (2023.12.0) and Python (3.12.1).

\begin{figure*}[t!]
\centering
\begin{minipage}{.346\textwidth}
  \centering
  \includegraphics[width=\linewidth]{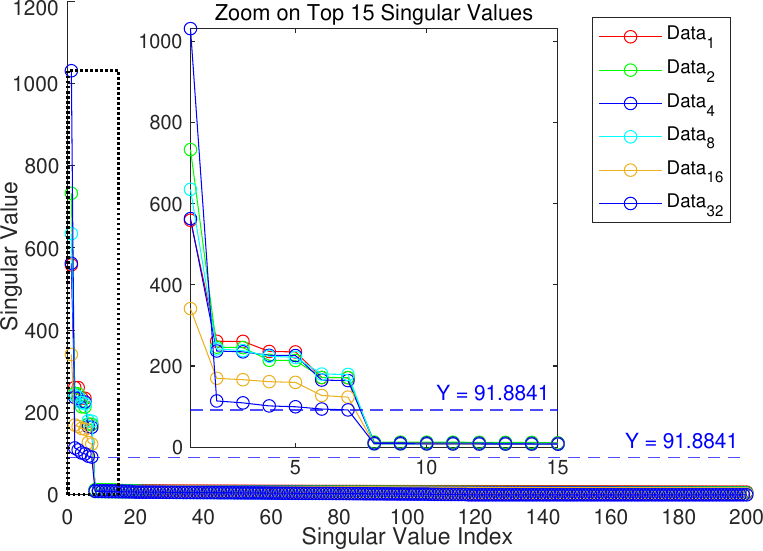}
  \caption{S-value distribution in various hardware levels.}
  \label{Singular Value1}
\end{minipage}%
\begin{minipage}{.306\textwidth}
  \centering
  \includegraphics[width=\linewidth]{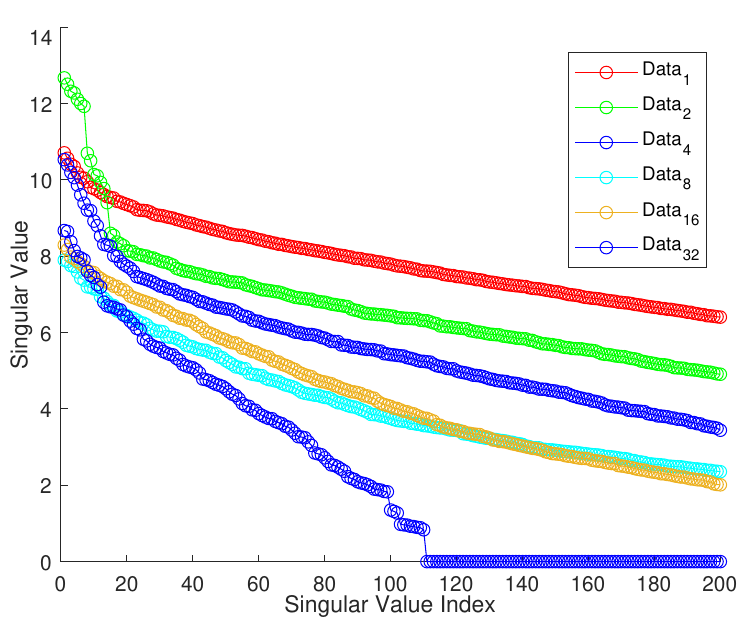}
  \caption{S-value of the reconstruction matrix.}
  \label{Singular Value2}
\end{minipage}%
\begin{minipage}{.348\textwidth}
  \centering
  \includegraphics[width=\linewidth]{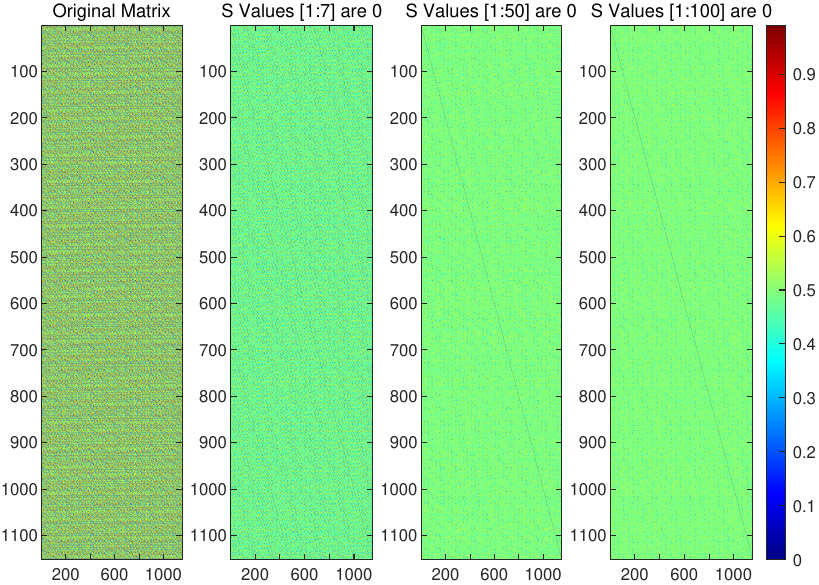}
  \caption{Extraction of different S-values.}
  \label{Bit Flips}
\end{minipage}
\label{fig:combined-figures}
\end{figure*}

\begin{figure*}[t!]
\centering
\begin{minipage}{0.25\textwidth}
  \centering
  \includegraphics[width=1\linewidth]{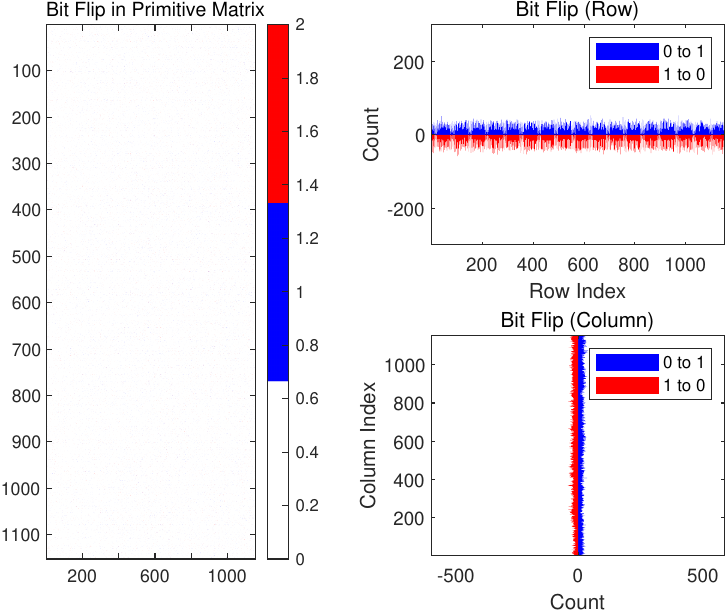}
  \caption{Raw in the same device.}
  \label{Correlation1.1}
\end{minipage}%
\begin{minipage}{0.25\textwidth}
  \centering
  \includegraphics[width=1\linewidth]{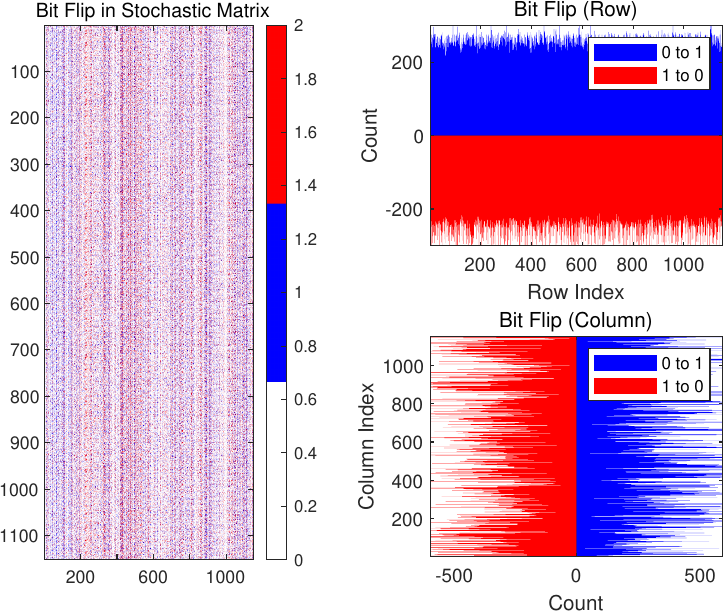}
  \caption{Optimising in the same device.}
  \label{Correlation1.2}
\end{minipage}%
\begin{minipage}{0.25\textwidth}
  \centering
  \includegraphics[width=1\linewidth]{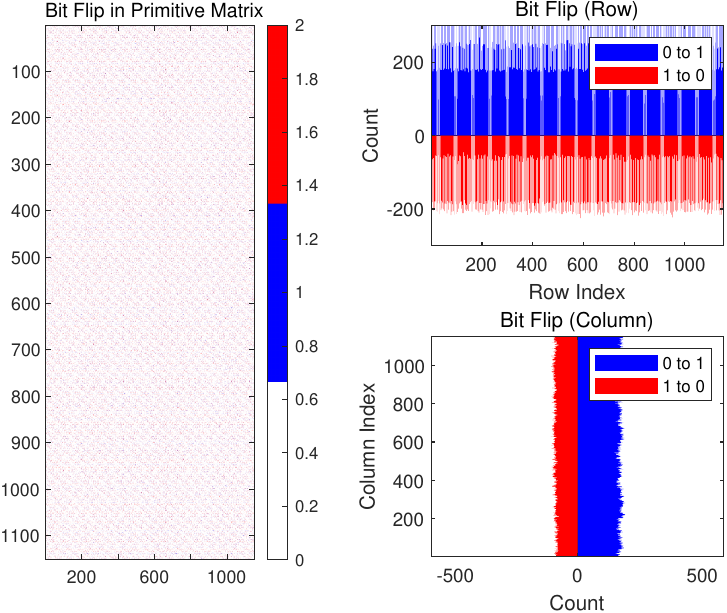}
  \caption{Raw in distinct devices.}
  \label{Correlation2.1}
\end{minipage}%
\begin{minipage}{0.25\textwidth}
  \centering
  \includegraphics[width=1\linewidth]{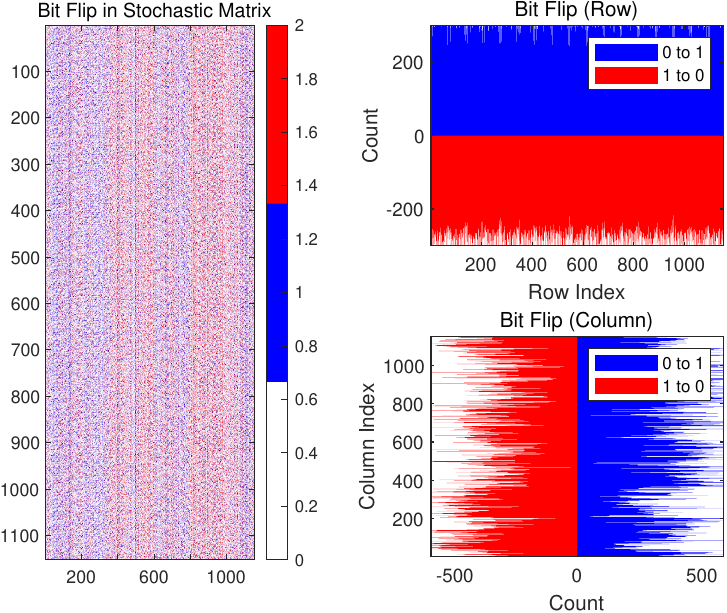}
  \caption{Optimising in distinct devices.}
  \label{Correlation2.2}
\end{minipage}
\label{fig:correlations}
\end{figure*}

\subsection{Pattern Analysis}
To demonstrate the theoretical feasibility of the proposed RO-SVD framework in practical hardware, we conducted simulations and analyses of frequency patterns and characterisation variations within multiple ring oscillators using the dataset \cite{github_reference}, which records 100 repeated measurements from 6592 ROs and counters deployed in 217 28nm Xilinx FPGAs and identical to the technology node we use. We visualised the characteristics of the ring oscillators in various placements (LEFT-UPPER, RIGHT-UPPER, LEFT-LOWER, and RIGHT-LOWER) within the layout, which provide complementary evidence to the findings of \cite{asha2021improving} that intra-die variation is spatially related and can be optimised based on slice placements, shown in Figure \ref{Pattern Analysis} (a). Furthermore, since the layout optimisation is based on the generation side by this dataset, this also proves that the intra-die variation with a rich set of intrinsic and stochastic patterns is characterised and optimised not only distributed across the hardware layouts but also available on the response generation side. Figure \ref{Pattern Analysis} (c) provides additional evidence of die-to-die variations in identical hardware series with the same technology node by setting up to visualise the various parameters within different devices, which is the foundation for using entropy sources as authentication applications in this work.

More importantly, we simulated the oscillatory process of ROs as affected by temperature, which proved to be intensely dependent on temperature. Specifically, by giving the measurements in the same device under the repetitions class and controlled temperature in Figure \ref{Pattern Analysis} (b), it can be easily observed that although the temperature is precisely controlled in the 1st, 50th and 100th repetitions, the frequency of the RO within the same device still contains a significant quantity of inherent patterns (white circular line) and stochastic patterns (black circular line) at the generation side of the response. This phenomenon shows that the generation side of the response contains a richer level of inherent and stochastic patterns (i.e., noise) in the natural environment where the temperature is hard to control precisely, which suggests that original responses generated by the entropy source can potentially be extracted and separated into intrinsic and stochastic patterns through downscaling and decomposition computing on the generation side for enhanced reconfigurability, which constitutes the theoretical basis for the implementation of this paper.

\subsection{Analysis of Singular Value Extraction}
Figure \ref{Singular Value1} shows the singular value distribution of the originally generated matrix across different entropy source hardware levels. It is evident that the first principal component of the matrix increases significantly as more entropy sources are introduced while the rank of the generated matrix decreases. This establishes the connection between the entropy sources and the first principal component of the matrix-based generation seed. Figure \ref{Singular Value2} shows the random matrix generated after extracting the first seven principal components with close and continuous singular values. This demonstrates that removing the first seven singular values for reconstruction is reasonable. We also compare the average Hamming distance within rows for different singular value removal cases, as shown in Figure \ref{Bit Flips}, which further justifies that our selected parameters are robust and customised for various content sizes.

\subsection{Matrix-based Authentication}
We constructed an authentication test using a $1024 \times 1024$ matrix to simulate the AI-generated images scenario with such size. The experiment demonstrates that our method can generate consistently invariant hash values by appropriately utilising the entropy source at the hardware level compared to existing non-matrix approaches under the same device conditions. This indicates that the proposed method effectively synthesises the principal components of the matrix and generates a principal component sequence for identifying similar entropy matrices produced by entropy source components from the same device. A demonstration of this process is shown in Figure \ref{Entropy Source}.

\subsection{Matrix-based Stochasticity Generation}
The generated random seeds are treated as a matrix, and a random matrix has a broad purpose in blockchain when elements are extracted row-by-row. To verify that the random sequence seeds of the original matrix generated by removing the principal components are different within the columns, we compare the average hamming distances between all rows to each other. For the same device case in Figure \ref{Correlation1.1}, we notice that without any matrices processing after continuous generation, the average proportion of the difference is 2.96\% shown in Figure \ref{Correlation1.2}. After applying our proposed method, the difference proportion increases to 43.80\%. For different devices case in Figure \ref{Correlation2.1}, our work increased the original proportion of differences from 18.88\% to nearly 50\% of the ideal value shown in Figure \ref{Correlation2.2}. This indicates that our design effectively mines stochasticity from the generation side. Furthermore, the original matrix tends to generate non-uniformly distributed 0s and 1s, which is not conducive to the quality of random numbers. The averaging step in the proposed method optimises this process and generates uniformly distributed 0s and 1s.

\begin{figure}[t!]
\centering
\includegraphics[width=1\linewidth]{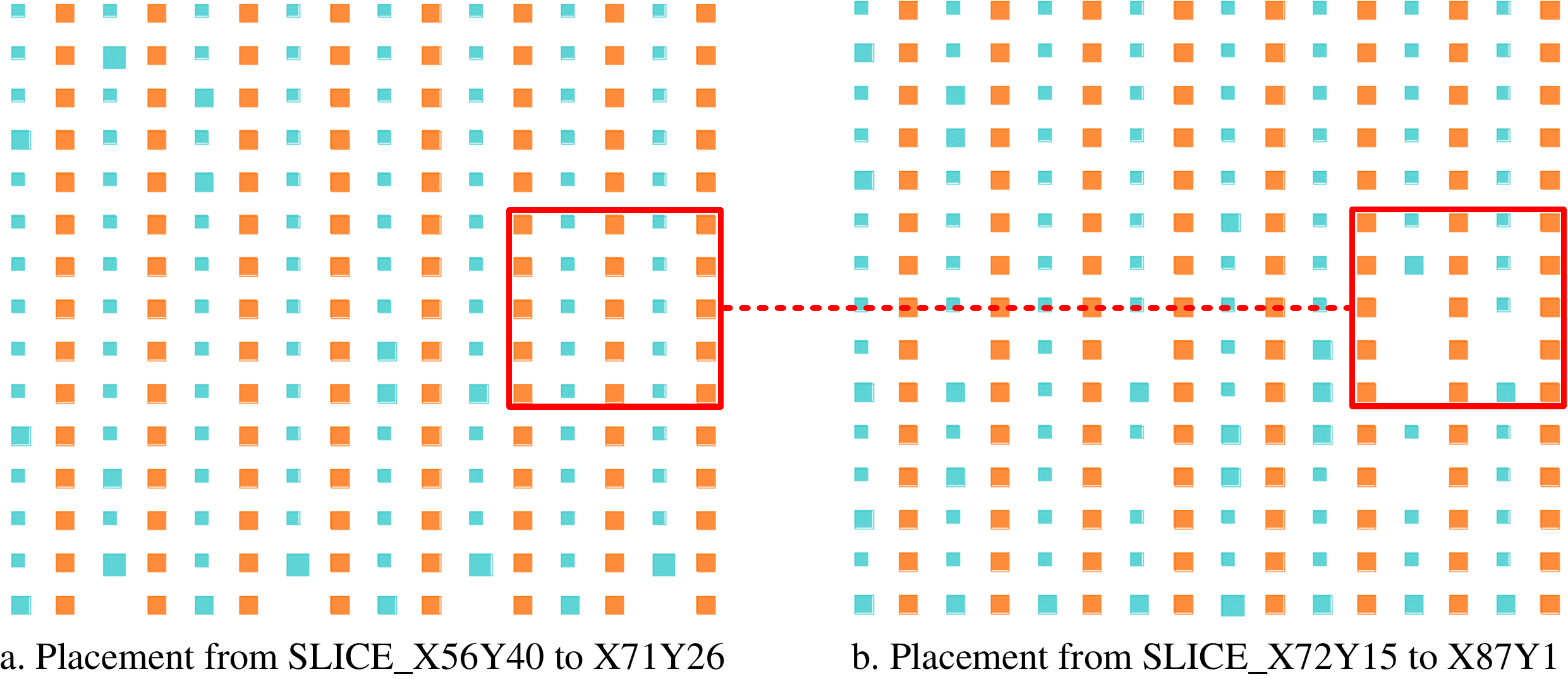}
\caption{Vivado Schematic Window with proposed constraint settings.}
\label{Placement}
\end{figure}

\subsection{Reconfigurability}
We enhance the reconfigurability of the design and significantly reduce development time by simplifying Xilinx Constraint Design, especially for cases requiring massive deployment for potential AIGC applications. ALLOW COMBINATORIAL LOOPS is a specific design constraint used to specify that the synthesis and timing analysis tools allow for combinational logic loops, which are the basis for ring oscillators to be implemented in FPGAs that can't be omitted. Fixed placement constraints LOC and BEL specify that look-up tables (LUTs) with their IOs are necessary for deploying entropy sources on hardware. Route-oriented physical constraints can reduce critical path variance to output stable responses. Timing constraints are crucial for entropy source design as they help avoid wasting timing optimisation resources on paths that do not affect functionality, focusing more on optimising critical paths to ensure the repeatability and stability of the entropy source output. Design constraints ensure consistent signal propagation delays, which are necessary for guaranteeing the repeatability and stability of the entropy source output in existing designs. A reasonable range of temperature and voltage can be specified by setting environmental constraints such as temperature and voltage, which ensure the entropy source produces a reliable and consistent response in different operating environments. Extensive constraints in the layout to generate stable and repeatable responses can reduce the reconfigurability of the design and increase development time. Typical constraints used for entropy source design in existing work compared to the proposed design are shown in Table \ref{Constraint}. Our design includes only essential constraints to enhance reconfigurability and reduce development time.

For AIGC applications, we utilise the reconfigurability of FPGAs to enable daily blockchain-based framework updates through a pre-registration process. The result of this reconfigurable strategy is visualised in Figure \ref{Placement}. By comparing the difference between the distributions in blue (non-fixed components, other) and yellow (fixed components, entropy sources) in Figure \ref{Placement} a and b, it is evident that when the fixed entropy sources change position by modifying the constraints, the position of the non-fixed surrounding components around them changes randomly and complex to predict thus introducing randomness by the parameters changes of the critical path. Additionally, randomness is introduced during the optimisation step of the layout routing phase of the EDA tool, resulting in different physical layouts of logic gates, flip-flops, routing, etc., even when re-synthesising an unchanged design. We can achieve low-cost renewal by identifying the updated principal components of entropy sources on the generation side.

\begin{table}[!t]
    \caption{Comparison of the Proposed and Existing Constraint Settings}
\label{Constraint}
\resizebox{\columnwidth}{!}{%
\begin{tabular}{|c|c|c|}
\hline
\textbf{Constraint Types} & \textbf{Existing Design} & \textbf{Proposed Design} \\ \hline
Compulsory Constraints\cite{mollinedo2022evaluation} & ALLOW COMBINATORIAL LOOPS & ALLOW COMBINATORIAL LOOPS \\ \hline
\multirow{3}{*}{Physical Constraints\cite{mahalat2021implementation}} & LOC, BEL & LOC, BEL \\ \cline{2-3} 
 & FIXED ROUTE & - \\ \cline{2-3} 
 & CREATE PBLOCK & - \\ \hline
\multirow{5}{*}{Timing Constraints\cite{tsiokanos2021dta}} & SET MULTI-CYCLE PATH & - \\ \cline{2-3} 
 & SET FALSE PATH & - \\ \cline{2-3} 
 & SET CLOCK GROUPS & - \\ \cline{2-3} 
 & SET MAX DELAY & - \\ \cline{2-3} 
 & SET MIN DELAY & - \\ \hline
Design Constraints\cite{plusquellic2022shift} & MAX FANOUT & - \\ \hline
Environmental Constraints\cite{huang2023design} & SET\_OPERATING\_COONDITIONS & - \\ \hline
\end{tabular}}
\end{table}

\subsection{Randomness Test}
The proposed design passes the National Institute of Standards and Technology (NIST) tests SP 800-22 Rev.1, which is a statistical test suite including 15 different official tests to ensure that the design does not exhibit any potential patterns before use. Furthermore, our generated seeds perform better than the original generation in terms of Rank, FFT and Nonoverlapping Template tests.

\subsection{Utilisation}
The summary of hardware costs is assumed to be based on the actual on-board implementation for AI-generated images with the size of $1024 \times 1024$ in this work. We used the Xilinx Power Estimator (XPE) suite to simulate and generate specific estimates of the total power consumption of the hardware intellectual property (IP) as a whole in the early design stage. 

\begin{table}[ht]
\caption{Utilisation of Hardware Cost and Power of Proposed RO-SVD}
\label{Utilisation}
\resizebox{\columnwidth}{!}{%
\begin{tabular}{|c|c|c|c|c|}
\hline
 & \begin{tabular}[c]{@{}c@{}}On-chips\\ LUT\end{tabular} & \begin{tabular}[c]{@{}c@{}}On-chips\\ Register\end{tabular} & \begin{tabular}[c]{@{}c@{}}On-chips\\ Delay\end{tabular} & \begin{tabular}[c]{@{}c@{}}On-chips\\ Power\end{tabular} \\ \hline
Entropy Source & 2600 & 512 & \multirow{3}{*}{0.943ns} & \multirow{3}{*}{1.24W} \\ \cline{1-3}
Buffer & 885 & 952 &  &  \\ \cline{1-3}
Others & 106 & 155 &  &  \\ \hline
SVD and Matrix Operations & 24667 & 27290 & 1.966ns & 3.85W \\ \hline
\end{tabular}%
}
\end{table}

\section{Conclusion and Future Work}
In this paper, we proposed a blockchain-based framework called ring oscillator-singular value decomposition (RO-SVD), which is the first hardware-based engineering study dedicated to discussing and implementing copyright traceability for AI-generated content and emphasises customisation, unpredictability, efficiency, management and reconfigurability. Based on comprehensive analysis and on-board experiments with multiple AI-applicable FPGAs, our hardware-software (HW/SW) co-design prototype shows good performance in randomness and enables work in parallel with existing FPGA-based AI acceleration devices as an added copyright traceability function. The proposed framework demonstrates the potential for efficient traceability, transactions and management of copyright information in blockchain for AI-generated content. Furthermore, our work inspires engineering solutions for emerging challenges in copyright issues of AI-generated content with multiple dimensions and various forms for robustness or non-robustness blockchain scenarios. For future work, we will aim to enhance the computing efficiency of each component to increase the speed of the overall design.

\section*{Acknowledgment}
This work is supported by Hong Kong Innovation and Technology Commission (InnoHK Project CIMDA), Hong Kong Research Grants Council (Project 11204821), and City University of Hong Kong (Project 9610460). 

We would like to thank the following researchers for their useful discussions: Yuhan She on framework construction, Gaoyu Mao on co-design, Abdurrashid Ibrahim Sanka on blockchains and Yao Liu on random number generation.

\bibliographystyle{IEEEtran}
\bibliography{RO_SVD_arXiv}
\end{document}